\documentclass[superscriptaddress,showpacs,A4,twocolumn]{revtex4}
\usepackage{psfig}
\usepackage{graphicx}
\usepackage[latin1]{inputenc}

\newcommand{\EVRY}{D\'epartement de Physique et Mod\'elisation,
Universit\'e d'Evry Val d'Essonne\\
Boulevard F. Mitterrand, 91025 Evry cedex}
\newcommand{\LKB}{Laboratoire Kastler Brossel, Universit\'e Pierre et Marie Curie\\
T12, Case 74, 4 place Jussieu, 75252 Paris, France}

\begin{document}
\title{Narrow-line phase-locked quantum cascade laser in the 9.2~micron range.}

\author{Franck Bielsa}
\affiliation{\LKB}
\affiliation{\EVRY}

\author{Albane Douillet}
\affiliation{\LKB}
\affiliation{\EVRY}

\author{Tristan Valenzuela}
\affiliation{\LKB}
\affiliation{\EVRY}

\author{Jean-Philippe Karr}
\email{hilico@spectro.jussieu.fr}
\affiliation{\LKB}
\affiliation{\EVRY}

\author{Laurent Hilico}
\email{hilico@spectro.jussieu.fr}
\affiliation{\LKB}
\affiliation{\EVRY}

\date{\today}
\begin{abstract}
We report on the operation of a 50~mW continuous wave
quantum cascade laser (QCL) in the 9.2~$\mu$m range, phase locked to a single mode CO$_2$ laser
with a tunable frequency offset.
The wide free-running emission spectrum of the QCL (3-5~MHz) is strongly narrowed down 
to the kHz range making it suitable for high resolution molecular spectroscopy.
\end{abstract}
%\ocis{120.3930,140.3470,999.9999 quantum cascade laser}
\maketitle

%%%%%%%%%%%%%%%%%%%%%%%%%%%%%%%%%%%%%%%%%%%%%%%%%%%%%%%%%%%%%%%%%%%%%%%%%%%%%%%%%%
%\section{Introduction}
Continuous wave high power ($\ge$50~mW) quantum cascade laser sources (QCL) recently became commercially available~\cite{QCL-continuous}.
They exhibit new features for infrared laser spectroscopy. Mid-IR QCLs are easily 
and widely tunable over more than 200~GHz 
by adjusting their temperature or injection current. Since they do not present phase-amplitude coupling, 
their ultimate linewidth is expected to be very narrow.
In practice, due to thermal instabilities, they 
present a wide free-running emission spectrum, in the MHz range~\cite{QCL-MHz-1,QCL-MHz-2}. 
Several experiments have shown a significant reduction of the QCL linewidth down
to the 10~kHz range by injection current locking to a molecular 
line~\cite{QCL-molec}, or well below the kHz range by locking to a Fabry Perot cavity~\cite{QCL-cav} 
using the Pound-Drever-Hall technique.
%%%%%%%%%%%%

%%%%%%%%%%%% 
Phase-locking is a well-known technique used to reduce the relative phase noise between 
two laser oscillators 
or to transfer the spectral features of a stable laser to a noisy one~\cite{santarelli}. This method consists in comparing the phase of the laser beat note with that of a RF synthesizer signal.
It has recently been applied to the frequency control of terahertz QCLs~\cite{betz}, 
but never so far to infrared QCL linewidth reduction. 

Our motivation for developping a frequency controlled QCL source is 
high resolution vibrational spectroscopy of the hydrogen molecular ions 
H$_2^+$ or HD$^+$. Indeed, 
those ions have recently been pointed out as good candidates for optical
proton to electron mass ratio determination~\cite{hilico,schiller,karr,roth-schiller}. 
Recent calculations have shown that two-photon vibrational spectroscopy of H$_2^+$ in the 
9.1-9.3~$\mu$m range is feasible~\cite{hilico} with interaction times of a few tens of ms. 
Transitions frequencies have been predicted with 1~ppb relative accuracy~\cite{korobov-relatH2,korobov}.
Further progress in QED correction calculations should allow proton to electron mass ratio
determination with a relative accuracy of 10$^{-10}$, in significant progress over 
the present one (CODATA 2002)~\cite{codata}. This corresponds to an uncertainty of a 
few kHz on the transition frequencies, hence the need for a kHz linewidth laser source.

In the 9.1-9.3~$\mu$m wavelength 
range, only two kinds of cw laser sources are available : CO$_2$ lasers and QCLs. 
Single mode CO$_2$ lasers can deliver up to several 
watt of optical power 
in a narrow bandwidth of less than 1~kHz. Unfortunately, their tunability covers a range of a few tens of MHz and the CO$_2$ emission lines do not overlap with the H$_2^+$ spectrum. 
In this paper, we show that it is possible to combine the advantages of both kinds of sources
by transferring the spectral properties of a CO$_2$ 
laser to a widely tunable QCL using 
phase-lock loop techniques. We demonstrate the operation of a tunable narrow-line high-power 
laser source suitable 
for two-photon spectroscopy of hydrogen molecular ions with a kHz resolution.
%%%%%%%%%%%%%%%%%%%%%%%%%%%%%%%%%%%%%%%%%%%%%%%%%%%%%%%%%%%%%%%%%%%%%%%%%%

\begin{figure}[h]
\center
\includegraphics[width=8cm]{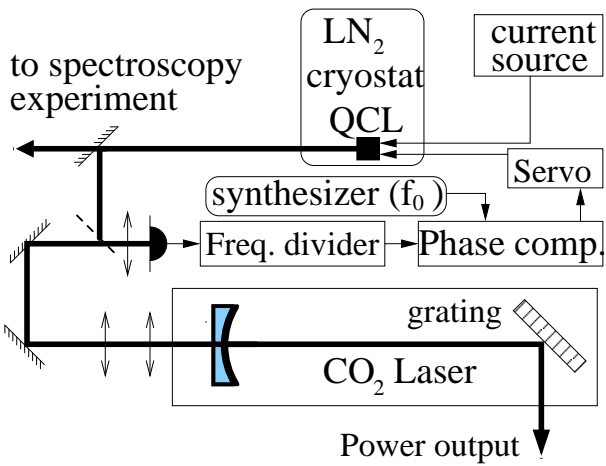}
\caption{\label{fig-setup} Simplified experimental setup. The low power CO$_2$ laser output  ($\approx$50~mW) is beated with a fraction of the QCL emission ($\approx$10~mW). The frequency divider and phase comparator are described in text, and the servo circuit in Fig.~\ref{servo}.}
\end{figure}

The experimental setup is depicted in Fig.~\ref{fig-setup}. 
The quantum cascade laser (SBCW496DN from AlpesLaser) is a single mode distributed feedback laser. 
When operated in cw regime under cryogenic conditions, it delivers up to 160~mW and is tunable between
9.16 and 9.24~$\mu$m (32.44 and 32.73~THz). Its threshold current is 400~mA at 77~K, and its maximum driving current is 1~A. The temperature 
and current 
tunabilities are 3~GHz/K and 150~MHz/mA.
%%%%%%%%%%% 
%%%%%%%%%%

%%%%%%%%%%
The QCL is mounted in a liquid nitrogen cryostat with a ZnSe output window and 
driven by a home-made, low-noise stabilized current 
source. The voltage across the QCL is about 9~V and the dynamic resistance is 1.8~$\Omega$. 
The electrical power dissipated in the QCL can be as high as 9 W, which requires efficient 
heat dissipation. 
For this purpose, the laser chip is fixed on a monolithic copper post, 
screwed on the cryostat cold plate in order to minimize 
the thermal resistance. The post can be heated and temperature controlled 
at the 10~mK level. 
With a 700~mA driving current, the QCL temperature stabilizes around 80~K without external 
heating and the QCL delivers an optical power of 50~mW at 9.166~$\mu$m, the required wavelength 
to probe the 
(v=0, L=2)$\rightarrow$(v=1,L=2) two-photon line in H$_2^+$.
%%%%%%%%%%

%%%%%%%%%%%
The output beam is collimated using ZnSe collimating optics from Cascade Technologies 
(CO-01, 0.85~mm N.A. and 1.6~mm working distance).
The far field transverse structure of the QCL beam presents 
a nodal structure 
with one main and two secondary lobes along the vertical axis. 
Along the horizontal axis, the Gaussian beam shape dependence on propagation distance is consistent with a M$^2$ parameter of about 3.3.
%%%%%%%%%%%%%%%%%%%%%%%%%%%%%%%%%  

%%%%%%%%%%%%%%
We use the CO$_2$ laser both as local
oscillator to characterize the QCL spectrum properties, and as stability reference to phase 
lock the QCL. It is a sealed-off, low pressure, dual-discharge-arm,  1~m long 
single longitudinal mode laser. The cavity is closed by a R$_{max}$ mirror at 9.2~$\mu$m and 
a 150~lines/mm grating in the Littrow configuration. The grating zero$^{th}$ diffraction order 
 is the main output of the laser (95\% efficiency).  
When operated with 13~Torr of standard gas mixture (CO$_2$-He-N$_2$) and 24~mA discharge current, 
the laser oscillation is obtained 
in the 9~$\mu$m band up to the 9R(48) line with more than 1~W of optical power. 
The  CO$_2$ laser emission spectrum has a linewidth in the kHz range~\cite{frech} and only very 
slow drifts (less than 1~MHz/s). 
%%%%%%%%%%%%%%%%%%%%%%%%%%%%%%%%%

\begin{figure}[h]
\center
\includegraphics[width=6cm,angle=-90]{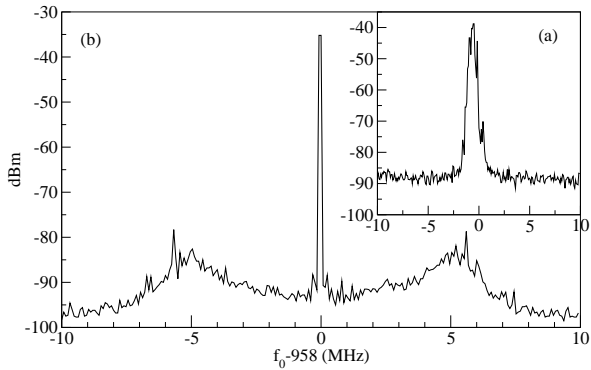}
\caption{\label{beat_notes} (a) Free-running and (b) phase-locked QCL/CO$_2$ beat note spectrum. (a) RBW 500~kHz, VBW 1~kHz, (b) RBW 10~kHz, VBW 1~kHz.}
\end{figure}

%%%%%%%%%%%%%%%%%%%
About 10~mW of QCL and 50~mW of CO$_2$ laser optical powers are mixed on a room temperature 
HgCdZnTe fast photodetector followed by a 37~dB low-noise RF amplifier. Although the beams' 
overlap is rather low, we obtain 
a signal to noise ratio of more than 45~dB in a 1~MHz resolution bandwidth for 
a beat frequency $f_0$ up to 1.5~GHz.  
The free-running QCL emission spectrum (Fig.~\ref{beat_notes}a) is about 3 to 5 MHz wide as already observed with 
other QCLs~\cite{QCL-MHz-1,QCL-MHz-2} and exhibits a low frequency jitter over more 
than 10~MHz.
%%%%%%%%%%%%%%%%%%%%%%%%%%%%%%%%%

%%%%%%%%%%%%%%%%%%
In order to efficiently phase lock the QCL on the CO$_2$ laser, we tailor a wideband phase error 
signal and use a standard second order feedback loop~\cite{gardner,blanchard,santarelli}.
The optical and electronic paths are shortened as much as possible (about 2~m) to minimize time delay.
The beat note signal is high-pass filtered above 700~MHz and divided by 8 using a MC12093 
high speed frequency divider.
The phase comparison with a synthesized reference signal is performed at $f_0$/8 using 
a MCH12140 phase/frequency detector
with a $\pm2\pi$ range. As a result, we obtain a very wide band ($\ge$10~MHz) $\pm16\pi$ phase/frequency 
comparison of the two lasers' spectra. The measured slope of the error signal is $s=0\cdot022$~V/rad.
%%%%%%%%%%

\begin{figure}[h]
\center
\includegraphics[width=8cm]{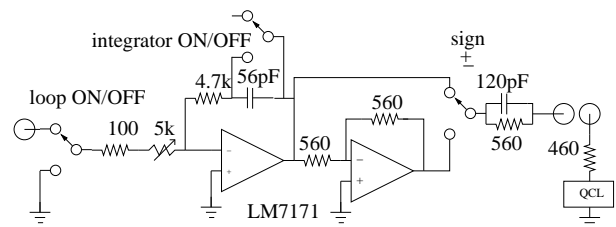}
\caption{\label{servo} Servo electronics of the phase lock loop.  
The resistor values are in $\Omega$.}
\end{figure}

%%%%%%%%%%%
The servo loop is depicted in Fig~\ref{servo}. It simply consists in an integrator with a 600~kHz cut-off frequency built with a fast operational amplifier. The feedback loop gain is adjusted 
using a variable resistor.
The correction signal is directly applied to the QCL through a $\approx$1k$\Omega$ resistor 
that limits the output current from the amplifier. This resistor is split into two parts. 
The first one belongs to a phase advanced filter with 2.4~MHz cut-off frequency adjusted to optimize the loop bandwidth. 
The second part is directly soldered on the QCL pads inside the cryostat in order to minimize wire capacitance effects.
Because we use a phase-frequency detector, we have to choose the suitable sign for the correction signal
depending on whether the QCL frequency is red or blue detuned with respect to the CO$_2$ laser frequency.
This is the aim of the second inversor follower amplifier.
%%%%%%%%%%%%%%%%%%%%%%%%%%%%%%%%%%%%%%%%%%%%%%%%%%%%%%%%%%%%%%%%%%%%%%
%%%%%%%%%%%%%%%

\begin{figure}[h]
\center
\includegraphics[width=8cm]{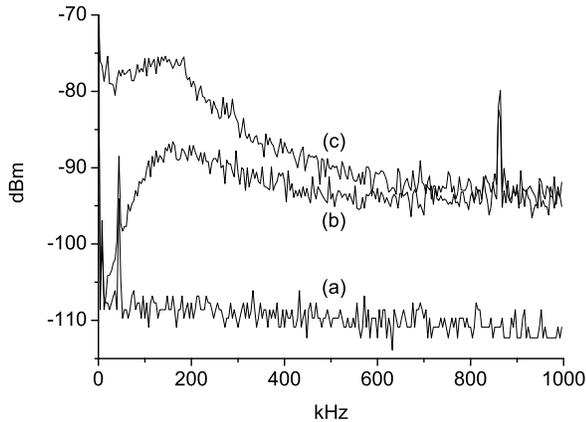}
\caption{\label{bf1M-1k-1k} Phase noise spectral density of the QCL/CO$_2$ beat note. 
(a) electronic noise, (b) proportional and integral corrections, (c) proportional correction only. 1~kHz RBW and VBW.}
\end{figure}

%%%%%%%%%%%%%%%
The beat note spectrum, taken with the tracking servo active, is shown in Fig.~\ref{beat_notes}b. It represents the relative phase noise spectral density between the QCL and CO$_2$ laser.
It exhibits an extremely narrow central peak with a -3~dB width of less than 200~Hz, limited by the 
resolution of the spectrum analyzer. This width is much smaller 
than the CO$_2$ laser one, meaning
that the CO$_2$ laser's spectral features are transferred to the QCL.   
The spectrum's wings show a servo-loop unity gain frequency
of the order of 6~MHz with a carrier 53~dB above the phase noise 
level in a 10~kHz resolution bandwidth. We now estimate the energy concentration ratio in the central peak. 
The normalized beat note spectrum can be expressed as~\cite{blanchard}
\begin{equation}
S(f)\approx e^{-\sigma_{\varphi}^2} \delta(f)+S_{\varphi}(f)
\end{equation}
where $\sigma_{\varphi}$ is the phase error variance, $\delta(f)$ the Dirac delta-function 
and $S_{\varphi}$ the phase noise spectral density. 
We have $S_{\varphi}(f)\approx 10^{-9.3}\approx5\cdot10^{-10}$rad$^2$/Hz.
Graphical integration within the loop bandwidth gives an estimation of half the actual phase variance, 
i.e. $\frac{1}{2}\sigma_{\varphi}^2=3\ 10^{-3}$~rad$^2$~\cite{gardner,blanchard}. 
Fig.~\ref{bf1M-1k-1k} shows the phase error spectral density at the output of the phase comparator. 
Comparison of curves (b) and (c) shows that the integrator reduces phase noise by more 
than 10~dB up to 200~kHz. The phase noise error signal spectral density exhibits a plateau 
at the S$_u$=10$^{-12.4}$~V$^2$/Hz level corresponding to a phase noise spectral density 
$S_{\varphi}=S_u/s^2=8\cdot2~10^{-10}$~rad$^2$/Hz. Taking into account the 6~MHz noise bandwidth, 
we obtain an alternative estimation of the phase variance $\frac{1}{2}\sigma_{\varphi}^2=0\cdot005$~rad$^2$
in good agreement we the first one. From those values, we can deduce that the energy concentration 
ratio in the narrow central peak $e^{-\sigma_{\varphi}^2}$ is higher than 99\%. 
Let us stress that the large feedback loop bandwidth is essential to obtain this result~\cite{note}.
%%%%%%%%%%%%%%%%%%%%%%%%%%%%%%%%%%%%%%%%%%%%%%%%%%%%%%%%%%%%%%%%
%%%%%%%%%%%%%%%%%%%%%%%%%%%%
%%%%%%%%%%%%%%%%%%%%%%%%%%%%%%%%
The phase locked operation of the QCL is easily obtained with a tunable frequency offset in 
the 200-1500~MHz range around the CO$_2$ emission line. The high frequency limit is due to the 
detector cut-off. The low frequency limit is the minimum beat frequency necessary 
to obtain a wide band 
error signal after division and phase comparison. It can be overcome by frequency shifting 
the high power output of the CO$_2$ laser with an acousto-optic modulator 
before beating with 
the QCL. Stable operation of the lock during several hours is obtained.
%%%%%%%%%%%%%%%%%%

In conclusion, we have shown that the spectral features of CO$_2$ lasers can be transferred to a QCL with a 
tunable frequency offset, making QCLs very interesting tools for high resolution molecular 
spectroscopy. We have used this source to observe several absorption lines
of the $\nu_6$ band of formic acid~\cite{hitran} 
in quasi-coincidence with the 9R36 to 9R42 emission lines of the CO$_2$ laser with a kHz 
resolution and measured their absolute frequencies~\cite{bielsa-HCOOH} to obtain suitable 
molecular frequency references for H$_2^+$ vibrational spectroscopy.
%%%%%%%%%%

We thank O. Acef, G. Santarelli and M. Lours (LNE-SYRTE), A. Vasannelli (Thales group) and Remy Battesti. 
Laboratoire Kastler Brossel is UMR 8552 du CNRS. This work was supported by an
ACI jeune 03-2-379 and BNM grants 02-3-008 and 04-3-009.
%%%%%%%%%%%%%%%%%%%%%%%%%%%%%%%%%%%%%%%%%%%%%%%%%%%%%%%%%%%%%%%%%%%%%%%%%%%

%%%%%%%%%%%%%%%%%%%%%%%%%%%%%%

\begin{thebibliography}{99}
\bibitem{QCL-continuous}A. Soibel, C. Gmachl, D. L. Sivco, M. L. Peabody, A. M. Sergent, A. Y. Cho, Appl. Phys. Lett. {\bf 83}, 24 (2003).
\bibitem{QCL-MHz-1}D. Weidmann, L. Joly, V. Parpillon, D. Courtois, Y. Bonetti, T. Aellen, M. Beck, J. Faist, D. Hofstetter, Optics Letters {\bf 28}, 704 (2003).
\bibitem{QCL-MHz-2}H. Ganser, B. Frech, A. Jentsch, M. Murtz, C. Gmachl, F. Capasso, D. L. Sivco, J. N. Baillargeon, A. L. Hutchinson, A. Y. Cho, W. Urban, Opt. Comm. {\bf 197}, 127 (2001).
\bibitem{QCL-molec}R.M. Williams, J.F. Kelly, J.S. Hartman, S. W. Sharpe, M. S. Taubman, J.L. Hall, F. Capasso, C. Gmachl, D.L. Sivco, J.N. Baillargeon, A.Y. Cho, Optics Letters {\bf 24}, 1844 (1999).
\bibitem{QCL-cav}M.S. Taubman, T. L. Myers, B. D. Cannon, R. M. Williams, F. Capasso, C. Gmachl, D.L. Sivco, A.Y. Cho, Optics Letters {\bf 27}, 2164 (2002), and references therein.
\bibitem{santarelli}G. Santarelli, A. Clairon, S.N. Lea, G.M. Tino, Opt. Comm. {\bf 104}, 339 (1994).
\bibitem{betz}A.L. Betz, R.T. Boreiko, B.S. Williams, S. Kumar, Q. Hu, J. L. Reno, Optics Letters {\bf 30}, 1837 (2005).
\bibitem{hilico}L. Hilico, N. Billy, B. Gr\'emaud, D. Delande, J. Phys. B {\bf 34}, 491 (2001).
\bibitem{schiller}S. Schiller, V.I. Korobov, Phys. Rev.A {\bf 71}, 032505 (2005).
\bibitem{karr}J-Ph. Karr, S. Kilic, L. Hilico, J. Phys. B {\bf 38}, 853 (2005).
\bibitem{roth-schiller}B. Roth, J. C. J. Koelemeij, H. Daerr, S. Schiller, Phys. Rev. A {\bf 74}, 040501(R) (2006).
\bibitem{korobov-relatH2}V.I. Korobov, Phys. Rev. A {\bf 74}, 052506 (2006).
\bibitem{korobov}V.I. Korobov, L. Hilico, J-Ph. Karr, Phys. Rev. A {\bf 74}, 040502(R) (2006).
\bibitem{codata}Review of Modern Physics {\bf 77}, 1 (2005).
\bibitem{frech}B. Frech, L.F. Constantin, A. Amy-Klein, O. Phavorin, C. Daussy, Ch. Chardonnet, M. M\"urtz, Applied Physics B {\bf 67}, 217 (1998).
\bibitem{gardner}F.M. Gardner, Phaselock technique (Ed. Wiley-intersciences, 1979).
\bibitem{blanchard}A. Blanchard, Phase-locked loops: applications to coherent receiver design, (Ed. Wiley, New York, 1976). 
\bibitem{note}Our first tries to phase lock the QCL were performed with a $\approx$600~kHz unitary loop 
gain frequency and provided a $\approx$10~rad$^2$ phase variance with no evidence for 
strong linewidth reduction.
\bibitem{hitran}L.S. Rothman, D. Jacquemart, A. Barbe, D. Chris Benner, 
M. Birk, L.R. Brown, M.R. Carleer, C. Chackerian Jr., K. Chance, L.H. Coudert, 
V. Dana, V.M. Devi, J.-M. Flaud, R.R. Gamache, A. Goldman, J.-M. Hartmann, 
K.W. Jucks, A.G. Maki, J.-Y. Mandin, S.T. Massie, J. Orphal, A. Perrin, 
C.P. Rinsland, M.A.H. Smith, J. Tennyson, R.N. Tolchenov, R.A. Toth, 
J. Wander Auwera, P. Varanasi, G. Wagner, 
J. Quant. Spec. Rad. Tr {\bf 96}, 139 (2005)
\bibitem{bielsa-HCOOH}F. Bielsa, K. Djerroud, A. Goncharov, A. Douillet, T. Valenzuela, C.
Daussy, L. Hilico, A.Amy-Klein, to be submitted to J. Mol. Spec.
\end{thebibliography}
\end{document}